\documentclass[reprint,5p,twocolumn]{elsarticle}
\usepackage{CJKutf8}
\usepackage{mathrsfs}
\usepackage{amssymb}
\usepackage{amsmath}
\usepackage{bbm}
\usepackage{graphicx}
\usepackage{bm}
\usepackage{float}
\usepackage[normalem]{ulem}
\usepackage{color}
\usepackage{multirow}
\usepackage{enumerate}
\usepackage{boondox-cal}
\usepackage{listings}
\usepackage{widetext}
\usepackage{url}

\usepackage{amssymb}

\usepackage{lineno}

\newcounter{bla}

\journal{Computer Physics Communications}

\begin{document}
	\begin{CJK}{UTF8}{gbsn}
		\begin{frontmatter}
			
			
			
			\title{OMR-NPA: Optimized Matrix Representation of Nucleon Pair Approximation}
			
			\author[b]{Y. Lei(雷杨)}
			\author[a]{Y. Lu(路毅)}
			
			\address[b]{School of National Defense Science and Technology, Southwest University of Science and Technology, Mianyang 621010, China}
			\address[a]{College of Physics and Engineering, Qufu Normal University, 57 Jingxuan West Road, Qufu, Shandong 273165, China}
			\begin{abstract}
				
				We optimize the matrix representation of the nucleon-pair approximation (NPA) of the nuclear shell model. The NPA is a widely adopted truncation approach of the nuclear shell model and proves to be effective in describing low-lying states of medium-heavy and heavy nuclei. Due to simplified (yet flexible) commutators and absolute elimination of angular momentum coupling, the matrix representation provides a formalism for the $M$-scheme NPA more efficient than others as far as we know. It also enables the practicable organization and storage design for intermediate results, including generated collective pairs, matrix products, and matrix traces, so that further optimization is achieved by reducing repetitive matrix operations, which are the most time-consuming procedures in the matrix-represented $M$-scheme NPA. We also describe optimizations specified for the $M$-scheme NPA, realized by invoking the Wigner-Eckart theorem, time-reversal symmetry, and conjugate operation of spherical tensors. Our optimization makes the combination of matrix representation and NPA more profitable. Such an implementation denoted by optimized matrix representation of NPA (OMR-NPA) is publicly released with open source. Its performance is analyzed and compared against unoptimized NPA codes.
			\end{abstract}
			
			
		\end{frontmatter}
		
		
		
		{\bf PROGRAM SUMMARY}
		
		\begin{small}
			\noindent
			{\em Program Title: OMR-NPA}                                          \\
			{\em Developer's repository link:} https://gitee.com/leiyang1985/OMR-NPA \\
			{\em Licensing provisions(please choose one):} GPLv3  \\
			{\em Programming language:} C/C++                                  \\
			{\em Supplementary material:}  OMR-NPA\_manual\_en.pdf                              \\
			{\em Nature of problem(approx. 50-250 words):}\\
			Many nuclear structural models adopt collective-pair configurations to construct trial wave-functions or the bases of the model space. It is generally difficult to calculate the overlap of such configurations, as well as their Hamiltonian matrix elements, which hinders further development of these pairing models and extension of their applicable region on the nuclear chart.
			\\
			{\em Solution method(approx. 50-250 words):}\\
			With the matrix representation of collective pairs, the overlaps of pair configurations, as well as their Hamiltonian matrix elements, can be always related to the traces of matrix products. The symmetries of pair structural matrices and their product trace enable an affordable way to store and retrieve intermediate matrix products so that some repetitive matrix operations can be avoided, and thus the computational time is reduced.

		\end{small}
		
		\section{Introduction}\label{int}
		Pair correlation universally exists in atomic nuclei. However, it's always the major issue to efficiently perform many-body calculations of nuclei with full consideration of nucleon pairing. Constrained by computing power back then, to serve a certain limited scientific merit, there are normally two approaches adopted to address this issue. Firstly, one can construct closed contraction relations of pair operators, by imposing sophisticated constraints on the collective pair structures. Then, the many-body calculation can benefit from the combination of like terms. Such an approach can be traced back to the seniority scheme \cite{sen-1, sen-2, sen-3} and Bardeen-Cooper-Schrieffer theory \cite{bcs}. However, the two-body degree of freedom is limited therein, and sometimes the particle number is no longer conserved. Secondly, one can emphasize the rotational symmetry of Hamiltonian, and impose angular momentum coupling to collective-pair configurations, which reduces the arbitrariness of collective-pair contraction and model space, while maintains the full two-body degrees of freedom and particle-number conservation. Nucleon-Pair Approximation (NPA) \cite{npa-rev-zhao} follows this philosophy. However, such an approach suffers unconfined collective-pair contraction and thus has difficulties to identify and make better use of intermediate results, e.g. repetitive pair commutators and configurations during calculations.
		
		It is desirable to develop a new method, which can make full use of repetitive intermediate results via dynamic programming, while takes full two-body degrees of freedom into account. Recently, the present author (Y. Lei) and collaborators developed a collective pair condensate variation method based on the matrix representation of collective pair structure \cite{pcv-lei}, which combines the above two considerations to some degree. However, such formalism enables only one type of collective pair in the many-body configuration, because it is specified for variational calculations. Later, the present authors and collaborator \cite{mnpa-lei} also noted that the matrix representation works for the $M$-scheme NPA \cite{mnpa-luo}. We suggested that the commutators between pair operators and one-body operators can be expressed as general matrix products, and thus a recursive formalism was obtained based on the trace calculations of these products. Such a formalism potentially provides higher computational efficiency with sophisticated storage and retrieval scheme of intermediate results.
		
		We apply the approach mentioned above to the NPA coding, denoted by OMR-NPA (Optimized Matrix Representation of the NPA), release it as open source attached with a brief code manual. In detail, we will describe how to efficiently identify, store and retrieve matrix products in practical NPA computations with the matrix representation. We benchmark the actual efficiency improvements with quantitative timing analysis of overlap calculations. 
		
		This paper is organized as follows:	In Sec.\ref{sec-for} we briefly describe basic formulas and briefly emphasize why NPA is much more efficient in $M$-scheme rather than traditional $J$-scheme; in Sec.\ref{sec-opt} we present optimization strategy for matrix operations; in Sec.\ref{sec-mem-poo} the storage of matrix products in a binary tree is recommended and carefully tuned; in Sec.\ref{sec-performance} we quantitatively analyze the improvement of our optimizations, based on timing comparison between overlap calculations before and after optimization; in Sec.\ref{sec-NPA-optimization} additional optimizations concerning rotational symmetry are presented; finally in Sec.\ref{sec-summary} we summarize this work.
		
		\section{basic concept}\label{sec-for}
		The NPA is a typical nuclear structural model, where the nuclear spectrum and electronic-magnetic properties are calculated with given Hamiltonian and model space. 
		
		Although the efficiency of the matrix-represented NPA has been benchmarked against $J$-scheme NPA in Ref.\cite{mnpa-lei}, we briefly summarize the advantages of the matrix-represented NPA related to the time consumption, since this paper aims to further improve the NPA efficiency. In the matrix-represented NPA, overlaps are coped with plain commutators, while in the $J$-scheme NPA, a series of intermediate quantum numbers is introduced in the recursive computations, which gives rise to substantially more complexity. The commutators in the matrix-represented NPA also can be straightforward with simply matrix operations, which has been extensively optimized in modern linear algebra libraries, like the MKL \cite{mkl}, while the $J$-scheme NPA commutators require angular momentum recoupling, and thus frequently involves the heavy computation of angular momentum algebra.
		
		As an optimized matrix-represented NPA, the crucial computation is to calculate the overlap of many-body configurations, which are constructed with several collective pairs. The comprehensive formalism about such computation is presented in Ref. \cite{mnpa-lei}. Here, we only highlight some essential concepts in our optimization.
		
		In the OMR-NPA, a collective pair is defined
		\begin{equation}\label{eq-pair-def}
			\begin{aligned}
				&\mathcal P^{\dagger}=\frac{1}{2}\sum_{ij}\mathcal p(ij) c^{\dagger}_ic^{\dagger}_j\\
				&\mathcal P=(\mathcal P^{\dagger})^{\dagger}=\frac{1}{2}\sum_{ij}\mathcal p(ij) c_jc_i,
			\end{aligned}
		\end{equation}
		where $c^{\dagger}$ and $c$ are orthogonal single-particle creation and annihilation operators, respectively. $\mathcal p(ij)$ is the pair structural coefficient, with anti-symmetry $\mathcal p(ij)=-\mathcal p(ji)$. Thus, all these $\mathcal p(ij)$ coefficients can be mapped onto a skew-symmetric matrix as
		\begin{equation}
			\mathcal p=
			\left[
			\begin{array}{cccc}
				0&\mathcal p(12)&\mathcal p(13)&\cdots\\
				-\mathcal p(12)&0&\mathcal p(23)&\cdots\\
				-\mathcal p(13)&-\mathcal p(23)&0&\cdots\\
				\cdots&\cdots&\cdots&\cdots
			\end{array}
			\right]\\
		\end{equation}
		
		In this work we use $\mathcal P$, $\mathbbm P$, $\Omega$, and $\Theta$ to represent the pair operators with the structural matrix  $\mathcal p$, $\mathbbm p$, $\omega$, and $\theta$, respectively.

		One-body operator in matrix representation is defined as
		\begin{equation}\label{eq-q}
			{Q} = \sum_{ij} q(ij) \hat{c}^{\dagger}_i \hat{c}_j,
		\end{equation}
		where coefficients $q(\alpha \beta)$ also constructs a matrix $q$ as			\begin{equation}\label{eq-q-coe-mat}
			q=
			\left[
			\begin{array}{ccc}
				q(11)&q(12)&\cdots\\
				q(21)&q(22)&\cdots\\
				\cdots&\cdots&\cdots
			\end{array}
			\right].
		\end{equation}
		Accordingly, the conjugate operator of $Q$, i.e., $Q^{\dagger}$, has its structural matrix as $q^\top$, i.e., the transpose matrix of $q$. In the NPA, some terms of Hamiltonian and transition operators are constructed with one-body operators.
		
		Here, we present three important contractions between collective pair and one-body operators. The contraction between two collective pairs reads
		\begin{equation}
			\left[\mathcal P,\mathbbm P^{\dagger}\right]=-\frac{1}{2}{\rm tr}\left(\mathbbm p\mathcal p\right)+Q,
		\end{equation}
		where ${\rm tr}(\mathbbm p\mathcal p)$ means the trace of $\mathbbm p\mathcal p$ matrix product, and the structural coefficient matrix of $Q$ is $q=\mathbbm p\mathcal p$. The contraction of a collective-pair operator, $\mathcal P$, and an one-body operator, $Q$, produces another collective-pair operator as
		\begin{equation}
			\mathbbm P=[\mathcal P,Q]~{\rm with}~
			\mathbbm p=\mathcal p q+q^\top\mathcal p,
		\end{equation}
		where $\mathbbm p$ is the structural coefficient matrix of $\mathbbm P$. The double contraction of one collective-pair creation operator, $\mathbbm P^{\dagger}$, and two collective-pair annihilation operators,$\mathcal P$ and $\Omega$, reads 
		\begin{equation}
			\Theta=\left[\mathcal P,\left[\Omega,\mathbbm P^{\dagger}\right]\right]~{\rm with}~
			\theta=\mathcal p\mathbbm p\omega+\omega\mathbbm p\mathcal p.
		\end{equation}
		Here, the property of the skew-symmetric matrix, $\mathbbm p^\top=-\mathbbm p$, is adopted.
		
		In odd-nucleon system, the unpaired nucleon is represented by single-particle operator. Arbitrary single-particle annihilation and creation operators read
		\begin{equation}
			\Phi=\sum_{i}\phi(i) c_i,~\Psi^{\dagger}=\sum_i\psi(i)  c^{\dagger}_i,
		\end{equation}
		where $\phi(i)$ and $\psi(i)$ are corresponding structural coefficients. In OMR-NPA, they are also mapped onto a row vector $\phi^\top$ and a column vector $\psi$, respectively, as
		\begin{equation}
			\Phi: \phi^\top=\left[
			\phi (1)~\phi(2)~\cdots
			\right] ~\Psi^{\dagger}: \psi=\left[
			\begin{array}{c}
				\psi(1)\\
				\psi(2)\\
				\cdots
			\end{array}
			\right].
		\end{equation}
		In the $M$-scheme NPA, the unpaired particle $\Psi^{\dagger}$ stays at a single spherical orbit without configuration mixing. Assuming such orbit is the $\alpha$-th orbit, then $\psi(i)=\delta_{i\alpha}$. Namely, all the vectors in the NPA calculation are one-hot corresponding to the spherical single-particle nature.
		
		The contraction of single-particle operators provides
		\begin{equation}
			\Phi\Psi^{\dagger}=\left[\Phi,\Psi^{\dagger}\right]-\Psi^{\dagger}\Phi=\phi\cdot\psi-Q,
		\end{equation}
		where $\phi\cdot\psi$ is the inner product of $\phi$ and $\psi$ vectors, and $Q$ is a one-body operator with $q=\psi^\top\phi$.
		
		Then, the pair configuration of a many-body system reads:
		the $M$-scheme NPA basis states is given by
		\begin{equation}\label{eqn:basis}
			{\mathcal P}^{\dagger}_0 {\mathcal P}^{\dagger}_1  \cdots {\mathcal P}^{\dagger}_N |0\rangle,
		\end{equation}
		where ${\mathcal P}^{\dagger}_0=1$ or $\Psi^{\dagger}$,  for even-or odd-nucleon system, respectively.
		
		The overlap of even-nucleon wave function will be frequently referred to in the NPA calculation. According to commutators provided above, the overlap can be calculated following the recursive relation as
		\begin{widetext}
			\begin{equation}\label{eq-ove-re}
				\begin{aligned}
					\langle 0| \mathcal P_1\mathcal P_2\cdots\mathcal P_N&|\mathbbm P^{\dagger}_1\mathbbm P^{\dagger}_2\cdots\mathbbm P^{\dagger}_N|0\rangle\\
					&=\sum_{k=1}^N -\frac{1}{2}{\rm tr}\left(\mathbbm p_N\mathcal p_k\right)\langle 0| \mathcal P_1\cdots\mathcal P_{k-1}\mathcal P_{k+1}\cdots\mathcal P_N|\mathbbm P^{\dagger}_1\cdots\mathbbm P^{\dagger}_{N-1
					}|0\rangle\\
					&+\sum_{k=2}^N\sum_{i=1}^k\langle 0|\mathcal P_1\cdots\mathcal P_{i-1}\mathcal P_{i+1}\cdots\mathcal P_{k-1}\mathcal P_{k+1}\cdots\mathcal P_N\Theta\{i,k\}|\mathbbm P^{\dagger}_1\cdots\mathbbm P^{\dagger}_{N-1
					}|0\rangle,\\
				\end{aligned}
			\end{equation}
		\end{widetext}
		where $\Theta\{i,k\}$ collective pair has structural coefficient matrix as $\theta\{i,k\}=\mathcal p_i\mathbbm p_N\mathcal p_k+\mathcal p_k\mathbbm p_N\mathcal p_i$.

		The overlap of an odd-mass basis is obtained in terms of the even-mass overlap and one-body operator matrix element, as below,
		\begin{equation}\label{eq-ove-o}
			\begin{aligned}
				\langle 0|\Phi  {\mathcal P}_1\cdots& {\mathcal P}_N|
				\Psi^\dagger {\mathbbm{P}}^{\dagger}_1 \cdots {\mathbbm{P}}^{\dagger}_{N}|0\rangle\\
				=& (\phi\cdot \psi) \langle  0| {\mathcal P}_1 \cdots {\mathcal P}_N
				| {\mathbbm{P}}^{\dagger}_1 \cdots {\mathbbm{P}}^{\dagger}_{N}|0\rangle\\
				&-\langle 0|  {\mathcal P}_1 \cdots {\mathcal P}_N |{Q}| \hat{\mathbbm{P}}^{\dagger}_1 \cdots \hat{\mathbbm{P}}^{\dagger}_{N} |0\rangle,
			\end{aligned}
		\end{equation}
		where ${{Q}}$ has structure matrix $ q = \psi\phi^\top$.
		
		Beyond Eqs. (\ref{eq-ove-re}) or (\ref{eq-ove-o}), all the matrix elements of arbitrary one-body operator $Q$, two-body operator $\Omega^{\dagger}\Omega$, and particle-hole type of interaction operator $QQ^{\dagger}$ are linear combinations of several even-mass overlaps. With standard procedure of many-body calculation, the nuclear level scheme and transition rate can be obtained. Corresponding formulas are already summarized in Ref. \cite{mnpa-lei}, and we don't repeat them here.
		
		\section{optimization of matrix operation}\label{sec-opt}
		According to Eq. (\ref{eq-ove-re}), the only large-scale floating-point operation in the overlap calculation is the matrix multiplication and trace calculation, which should be the most time-consuming computation in the OMR-NPA calculation. Many of them can be avoided according to the symmetry (or anti-symmetry) of the structural coefficient matrix. In this section, we will present the detail of our optimization following this philosophy.
		
		\subsection{encoding of matrix product}\label{sec-rep}
		The OMR-NPA calculation involves two types of pair structural matrices and single-particle structural vectors. Firstly, there are several collective pairs and unpaired particles in the $M$-scheme NPA basis, Hamiltonian, and one-body operators, whose structures are required to be manually inputted at the beginning of the calculation. We assign a specified integer to represent each inputted pair or unpaired particles, as well as its structural matrix/vector. Secondly, there are massive collective pairs and unpaired particles generated from the commutations described in Sec. \ref{sec-for}, e.g. the $\Theta\{i,k\}$ pair in Eq. (\ref{eq-ove-re}). According to the contraction formalism, the generated structural coefficient matrices and vectors are always the linear combinations of the products of manually inputted structural matrices or vectors (the first type of matrices and vectors). Since only a limited number of matrices and vectors are inputted in the NPA, there exist massive repetitive matrix products, which leaves room for further optimization.
		
		To identify these repetitive matrix products, we suggest creating an encoding rule to express every matrix product with simple unsigned char arrays. Then, the identification problem becomes a string matching problem, which already has various solutions. The encoding rule can be flexible depending on the coder's own preferences. Here is our choice. We use an array as ``\{{\it $i_1$,$n_1$,$i_2$,$n_2$,$i_3$,$n_3$,$\cdots$}\}'' to represent the matrix product as $\mathcal p_{i_1}^{n_1}\mathcal p_{i_2}^{n_2}\mathcal p_{i_3}^{n_3}\cdots$, where $\mathcal p_{i_1}$, $\mathcal p_{i_2}$, $\mathcal p_{i_3}$ $\cdots$ are the structural matrices or vectors manually inputted at the beginning of the NPA calculation with assigned integers $i_1$, $i_2$, $i_3$ $\cdots$ to specify them, and $\mathcal p^n$ means the $n$-th power of matrix $\mathcal p$. For example, the array of ``\{{\it i,1,N,1,k,1}\}'' corresponds to the $\mathcal p_i\mathbbm p_N\mathcal p_k$ product for the newly generated pair $\Theta\{i,k\}$ in Eq. (\ref{eq-ove-re}). 
		
		Such unsigned char arrays can be adopted as key values to store and retrieve the matrix elements of calculated matrix products with string matching, so that repetitive matrix multiplications can be avoided. We take the calculation of $\Theta\{i,k\}$ in (\ref{eq-ove-re}) as an example, where $\Theta\{i,k\}$ has structural matrix as $\theta\{i,k\}=\mathcal p_i\mathbbm p_N\mathcal p_k+\mathcal p_k\mathbbm p_N\mathcal p_i$. Assuming we just begin the computation, and no result of matrix multiplication is stored, we should first perform the actual matrix multiplication of $\mathcal p_i\mathbbm p_N\mathcal p_k$, store the resultant matrix elements in the memory pool orderly, and label its key as ``\{{\it i,1,N,1,k,1}\}''. Then we also need $\mathcal p_k\mathbbm p_N\mathcal p_i$, whose calculation, however, can be omitted, because it is encoded as ``\{{\it k,1,N,1,i,1}\}'', and one can easily identify the transpose of this matrix product represented with ``\{{\it i,1,N,1,k,1}\}'' already exists in the key set of our memory pool, given $\mathcal p_k\mathbbm p_N\mathcal p_i=-(\mathcal p_i\mathbbm p_N\mathcal p_k)^\top$. Thus, a simple transpose operation of stored $\mathcal p_i\mathbbm p_N\mathcal p_k$ matrix product provides the desired result. One sees that the above approach corresponds to an exchange of time complexity and space complexity. It works for a time-sensitive task, like the NPA calculation itself.
		
		With the array encoding,  one can easily identify the symmetric part in a matrix product before the actual multiplication, which enables optimized APIs (Application Programming Interface) specified for symmetric matrix under the BLAS standard \cite{blas}. For example, in the matrix produced expressed as ``\{{\it i,3,j,2,i,3,k,3}\}'', corresponding to $ \mathcal p_i^3 \mathcal p_j^2\mathcal p_i^3\mathcal p_k^3$, the longest symmetric part is identified as ``\{{\it i,3,j,2,i,3}\}'', given $ (\mathcal p_i^3 \mathcal p_j^2\mathcal p_i^3)^\top= \mathcal p_i^3 \mathcal p_j^2\mathcal p_i^3$. So we obtain the  $\mathcal p_i^3 \mathcal p_j^2\mathcal p_i^3$ product firstly and then use the symmetric matrix multiplication routine to calculate $ \mathcal p_i^3 \mathcal p_j^2\mathcal p_i^3\times \mathcal p_k^3$. Otherwise, if we use the array of matrix elements to identify the symmetric matrix in a matrix product, such a pre-identification can not be realized, and one may choose to calculate the product sequentially and lose one opportunity to use the symmetric matrix multiplication routine. 
		
		\subsection{encoding of pair}
		As stated in Sec. \ref{sec-rep}, there are two types of collective pairs involved in the OMR-NPA calculation: the manually inputted ones and generated ones from commutations. Supposedly, it is difficult to organize the generated collective pairs, and thus to optimize the matrix operations related to them. Since we already encode the matrix products with the unsigned-char arrays, the collective pair could readily be expressed following the same convention, which the NPA calculation can benefit from.
		
		We also take $\Theta\{i,k\}$ in Eq. (\ref{eq-ove-re}) as an example. Its structural matrix is $\theta\{i,k\}=\mathcal p_i\mathbbm p_N\mathcal p_k+\mathcal p_k\mathbbm p_N\mathcal p_i$, which is an linear combination of two matrix products, encoded as ``\{{\it i,1,N,1,k,1,}\}'' and ``\{{\it k,1,N,1,i,1}\}'' arrays, respectively. Then $\Theta\{i,k\}$ can be represented with concatenation of such two arrays as ``\{{\it 1,i,1,N,1,k,1,255,1,k,1,N,1,i,1}\}'', where ``255'', the largest integer that can be expressed with unsigned char, is adopted as the separation between two matrix product, the first and fifth ``1''s are the combination coefficients, and the rest elements are directly from original arrays. If $\mathcal p_i=\mathcal p_k$ then, the $\Theta\{i,k\}$ structural matrix is reduced to $2\mathcal p_i\mathbbm p_N\mathcal p_k$, and corresponding array would be ``\{{\it 2,i,1,N,1,k,1}\}''. If we further require $\mathcal p_i=\mathcal p_k=\mathbbm p_N$, a simply third power of $\mathcal p_i$ is obtained as $2\mathcal p_i^3$ with the array as ``\{{\it 2,i,3}\}''. This generated pair will be involved in another double commutation of further recursive overlap calculations, e.g., $[\mathcal P_j,[\Theta\{i,k\},\mathbbm P_{N-1}^\dag]]$, and generate another pair with structural coefficient matrix $2\mathcal p_i^3\mathbbm p_{N-1}\mathcal p_j+2\mathcal p_j\mathbbm p_{N-1}\mathcal p_i^3$ with array of ``\{{\it 2,i,3,N-1,1,j,1,255,2,j,1,N-1,1,i,3}\}''. Thus, all the collective pairs involved in our calculation can be encoded in an unified convention. 
		
		With the pair encoding, we can further reduce the number of recursions. We note that there can be many identical generated collective pairs in a single overlap calculation, even if they have different contraction history. Those identical pairs may introduce repetitive recursion calls. With the unsigned-char array to represent collective pair structure, we can more easily identify repetitive recursions and combine like terms accordingly. Then, the number of recursions and computational time further decrease. For example, the overlap calculation of $\langle\mathcal P_1\mathcal P_2\mathcal P_1|(\mathbbm P^{\dagger}_3)^3\rangle$ recursively requires two sub-overlaps as $\langle[\mathcal P_1,[\mathcal P_2,\mathbbm P^{\dagger}_3]]\mathcal P_1|(\mathbbm P^{\dagger}_3)^2\rangle$ and $\langle\mathcal P_1[\mathcal P_2,[\mathcal P_1,\mathbbm P^{\dagger}_3]]|(\mathbbm P^{\dagger}_3)^2\rangle$, according to Eq. (\ref{eq-ove-re}). With the first glance, one may feel that these two sub-overlaps involves two different pairs from the different double contractions as $[\mathcal P_2,[\mathcal P_1,\mathbbm P^{\dagger}_3]]$ and $[\mathcal P_1,[\mathcal P_2,\mathbbm P^{\dagger}_3]]$. However, if we express the structural coefficient matrices of these two pairs with unsigned-char arrays, one sees that they are actually identical pair expressed as ``\{{\it 1,1,1,3,1,2,1,255,1,2,1,3,1,1,1}\}''. Thus, these two sub-overlaps equal to each other, and we only need to recursively calculate one of them.
		
		The pair encoding also saves the memory for matrix expression. Conventionally, the expression of one matrix is an array of its all elements by double-float type, and thus requires $8N^2$-byte space, where $N$ is the number of involved single-particle orbits, and $N\geq 20$ for usual NPA calculations. If such a matrix is represented by an unsigned-char array, it only requires few bytes.
		
		\subsection{trace calculation}\label{sec-tr}
		In the OMR-NPA, all the matrix products finally are introduced to a trace calculation as shown in Eq. (\ref{eq-ove-re}). We note that most of the matrices inputted in the NPA are symmetric or anti-symmetric, and all the vectors are one-hot, which enables three optimizations for trace calculation, as follows.
		
		Firstly, we can reduce the trace calculations, under the general symmetries of
		\begin{equation}\label{eq-sys-tr}
			{\rm tr}(A)={\rm tr}(A^\top),~{\rm tr}(AB)={\rm tr}(BA),
		\end{equation}
		where $A$ and $B$ are general square matrices. Since some matrices are symmetric or anti-symmetric in the NPA, the traces of matrix products with the same set of matrices but different orders sometimes have the same value. For example, given $\mathcal p_1$, $\mathcal p_2$, $\mathcal p_3$, and $\mathcal p_4$ are all anti-symmetric matrices,
		\begin{equation}
			\begin{aligned}
				{\rm tr}(\mathcal p_1\mathcal p_2\mathcal p_3\mathcal p_4)&={\rm tr}(\mathcal p_2\mathcal p_3\mathcal p_4\mathcal p_1)\\
				&={\rm tr}(\mathcal p_3\mathcal p_4\mathcal p_1\mathcal p_2)\\
				&={\rm tr}(\mathcal p_4\mathcal p_1\mathcal p_2\mathcal p_3)\\
				={\rm tr}(\mathcal p_4\mathcal p_3\mathcal p_2\mathcal p_1)&={\rm tr}(\mathcal p_3\mathcal p_2\mathcal p_1\mathcal p_4)\\
				&={\rm tr}(\mathcal p_2\mathcal p_1\mathcal p_4\mathcal p_3)\\
				&={\rm tr}(\mathcal p_1\mathcal p_4\mathcal p_3\mathcal p_2)
			\end{aligned}
		\end{equation}
		We only need to calculate one of them, and the calculations for others can be omitted.
		
		Secondly, if there is symmetric or skew-symmetric matrix in a certain matrix product, its trace calculation can be converted into an inner product of two vectors, which requires less computation. Assuming $A^\top=\pm A$, and $B$ is another general square matrix, the trace of their product is calculated as
		\begin{equation}\label{eq-mat-dot-en}
			{\rm tr}(AB)=\sum_{ij}A_{ij}B_{ji}=\pm\sum_{ij}A_{ji}B_{ji}
		\end{equation}
		If one adopts the BLAS standard for the $A$ and $B$ matrix storage, i.e., the $A$ and $B$ matrices are stored as a one-dimensional vector with $n^2$ elements, then the above trace calculation is equivalent to the inner product of two vectors, which has only $O(n^2)$ time complexity, instead of $O(n^3)$ for matrix multiplication. Therefore, we should always shift a symmetric matrix or anti-symmetric one to the beginning of the matrix product before its trace calculation, according to the symmetry of Eq. (\ref{eq-sys-tr}). Shifting an anti-symmetric matrix is preferable because the symmetric part should be maintained for symmetric matrix multiplication routine, which can more effectively reduce computational time.
		
		Thirdly, for the odd-nucleon system, structural vectors are introduced to represent unpaired particles and complicates the trace computation. However, since all the structural vectors are hot-one as stated in Sec. \ref{sec-for}, the computational complication brought by them can be eliminated according to the hot-one property as described follows. According to Eq. (\ref{eq-ove-o}), all the single-particle operators are packed into a one-body operator for the odd-mass overlap calculation, and thus the two vectors of both unpaired particles from bra and ket will be maintained as a whole throughout the NPA calculation. That suggests the trace calculation for the odd-nucleon system may include either one row vector and one column vector, or no vector at all. For traces with vectors, we can always reorganize them in the form of ${\rm tr}(\phi^\top\prod_{s}\mathcal p^{n_s}_s\psi)$ according to Eq. (\ref{eq-sys-tr}), where $\phi^\top$ and $\psi$ are one-hot row and column vectors relocated at the beginning and end of the matrix-vector product, respectively. Supposing $\phi(i)=\delta_{i\alpha}$ and $\psi(i)=\delta_{i\beta}$, ${\rm tr}(\phi^\top\prod_{s}\mathcal p^{n_s}_s\psi)=\prod_{s}\mathcal p^{n_s}_s(\alpha\beta)$, which is the $(\alpha,\beta)$ element of $\prod_{s}\mathcal p^{n_s}_s$ matrix product. Namely, the $\prod_{s}\mathcal p^{n_s}_s$ matrix product can be only calculated once, and the trace values for all the single-particle scenarios with different $(\alpha,\beta)$ indexes are obtained. The computational time supposedly is reduced. Furthermore, with this approach, all the matrix-vector multiplications in the odd-nucleon system can be converted to matrix-matrix multiplications, which are already well-optimized herein. Therefore, we do not recommend further matrix-vector optimization, which may complicate the implementation.
		
		\section{matrix storage management}\label{sec-mem-poo}
		In the OMR-NPA, we selectively and orderly store and retrieve some matrix products in the memory to avoid repetitive matrix multiplications, as described in Sec. \ref{sec-rep}. In this section, we describe the matrix storage in detail.
		
		We chose a classic binary tree data structure to build up our matrix-product storage, over static storage, dynamic array, and balance tree data structure.  Static storage with continuous memory space can be the most efficient for data retrieval. However, it requires the determination of which matrix products will be most frequently used, and thus deserve to be stored before the actual calculation. Such a pre-determination is very difficult, chaotic, and thus also inapplicable for complicated computation like the NPA. With the dynamic array, one does not need the pre-determination, but the NPA calculation will create massive matrix products and insert them into continuous memory storage, which involves large-scale memory shifting. The time consumption for such dynamical memory shifting is unacceptable as evidenced by test calculation. The balance binary tree structure is for dynamic storage without memory shifting, and thus supposedly more efficient than continuous storage. However, to maintain the tree balance, large-scale rotation operations are still required, and its time consumption is comparable to that of dynamic memory shifting in the continuous memory storage as we have tested with red-black tree structure. The classic tree structure could perfectly avoid the memory shifting and rotation, and also enables relatively efficient retrieval with the bisection. Thus, it becomes our final choice.
		
		In the binary tree, every node includes a key for retrieval, and a value to point to the storage. More specifically, in the OMR-NPA, the key is defined as the unsigned-char array for a certain matrix product, while the value is defined as the memory address of corresponding matrix-element storage. As described in Sec. \ref{sec-rep}, whenever a matrix product is required, we encode it into a key with the encoding convention as mentioned in Sec. \ref{sec-rep}, and then search for such a key along the binary tree. If its key or its transpose ' key exists already, one can obtain the elements of matrix-product directly from the storage, and further multiplication is avoided; otherwise, we calculate the required matrix product and create a node on the tree for it.
		
		Because of the unconfined collective-pair contraction, it is usually unrealistic to store all the calculated matrix products. We have to give up the storage for some products, which is less frequently involved. Empirically, if a product has more matrices, it is less frequently to be involved. More matrices in the product mean a larger length of the unsigned-char array to represent it. For example, the product of $\mathcal p^{n_1}_1\mathcal p^{n_2}_2$ represented by ``\{{\it 1,$n_1$,2,$n_2$}\}'' with array length of 4 should be more frequently involved than that of $\mathcal p^{n_1}_1\mathcal p^{n_2}_2\mathcal p^{n_3}_3$ by ``\{{\it 1,$n_1$,2,$n_2$,3,$n_3$}\}'' with length of 6. Correspondingly, the latter has a longer representative array. Therefore, we take the length of the unsigned-char array as the measure to determine whether or not to store the corresponding matrix product.
		
\begin{figure}
			\includegraphics[angle=0,width=0.45\textwidth]{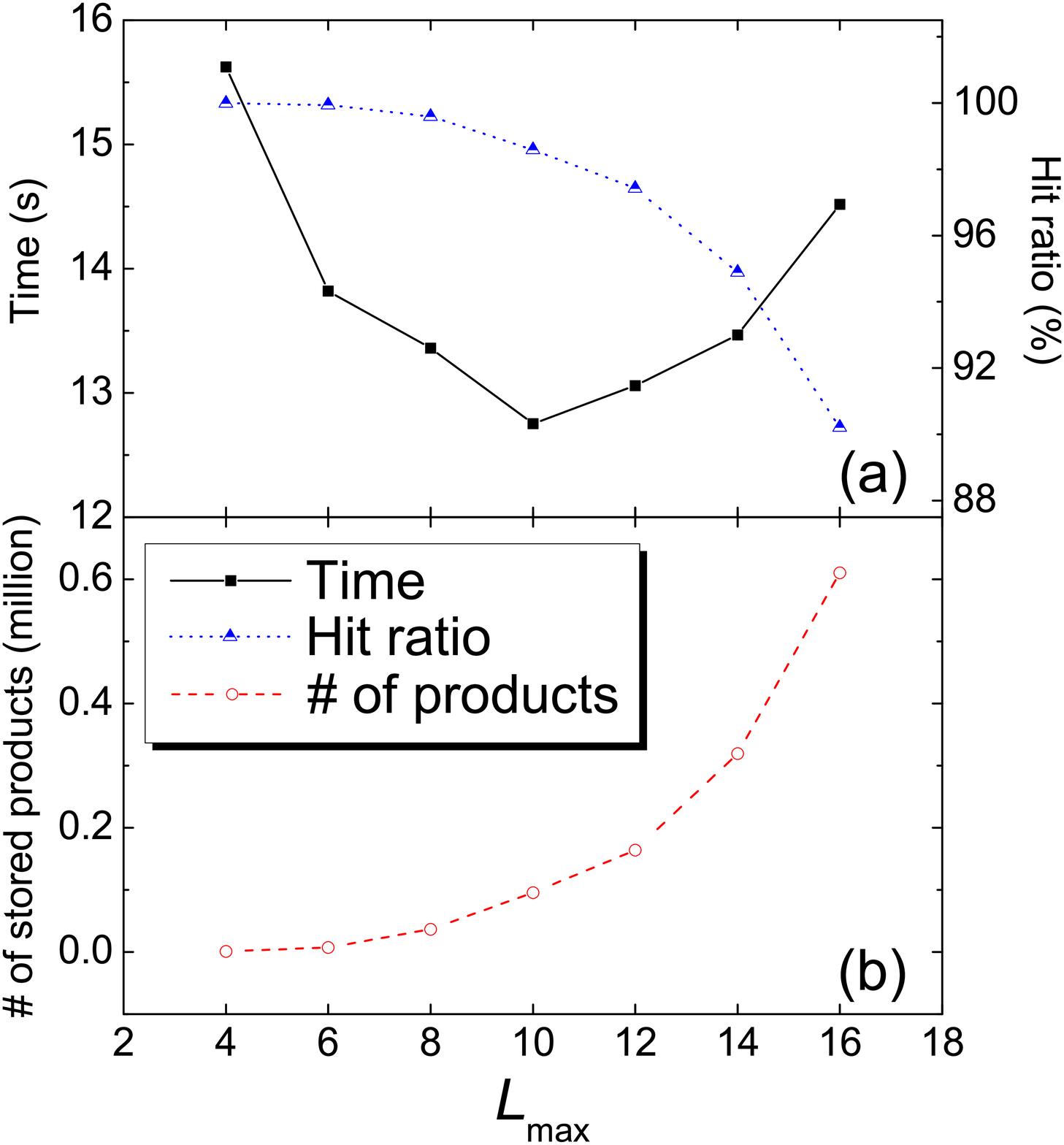}
			\caption{(color online) Average computational time for one overlap, hit rate of the matrix-product storage, and stored number of matrix products against $L_{\rm max}$ in the overlap calculation of the $SD$-truncated NPA $M$-scheme base in $pf$ shell with 12 nucleons. (see Ref. \cite{mnpa-lei} for more detail about the $M$-scheme NPA). $L_{max} = 10$ is efficiently optimal. However, due to limitation of memory footprint, we are often restricted to $L_{max} = 8$. (see the text for more details)}\label{fig-len-tun-en}
		\end{figure}

		We set an upper limit of storage for the array length of the matrix product. Such a limit is denoted by $L_{\rm max}$. For example, if $L_{\rm max}=4$, then we store the matrix elements of $\mathcal p^{n_1}_1\mathcal p^{n_2}_2$, while neglect $\mathcal p^{n_1}_1\mathcal p^{n_2}_2\mathcal p^{n_3}_3$. In order to get the optimal $L_{\rm max}$, we calculate all the overlaps between the $SD$-truncated NPA $M$-scheme base in $pf$ shell with 12 nucleons. The $SD$ truncation utilizes only $SD$ pairs with $L^{\pi}=0^+$ and $L^{\pi}=2^+$ to construct the model space. In Fig. \ref{fig-len-tun-en}, we present the computational time, the hit ratio of the storage, and the number of stored matrix products against $L_{\rm max}$, where the hit ratio is defined as
		\begin{equation}
			{\rm Hit ~ratio}=\frac{\rm count~ of~ successful~ retrieval}{\rm total ~count ~of ~retrieval}
		\end{equation}
		
		In Fig. \ref{fig-len-tun-en}(a), the computational time is a decreasing function of $L_{\rm max}$ for $L_{\rm max}\leq 10$, which suggests that the storage of matrix products indeed effectively reduces the matrix multiplication. After $L_{\rm max}= 10$, the number of storage matrices expands dramatically, as shown in Fig. \ref{fig-len-tun-en}(b), which reverses the tendency of the retrieval time. Thus, from a time-consuming point of view, $L_{\rm max}= 10$ may be optimal. According to the hit ratio therein, the smaller $L_{\rm max}$ leads to a higher hit ratio. This is because smaller $L_{\rm max}$ only allows us to store ``shorter" and thus more frequently involved matrix products, which will increase the chance of successful retrieval. For $L_{\rm max}\leq 8$, the hit ratio is always near 100\%, and thus $L_{\rm max}\leq 8$ cases all can be considered to be hit-ratio optimal. According to the storage magnitude in Fig. \ref{fig-len-tun-en}(b), the number of storage matrix products with with $L_{\rm max}= 10$ is about 100,000. If the OMR-NPA calculation is performed in the 82-126 major shell, where the structural coefficient matrix dimension is 44, then the double-precision memory storage is up to $\sim$1.5GB. If the Hamiltonian calculations are further introduced, the memory footprint could increase by at least 1 magnitude up to 10GB. To ensure that the OMR-NPA computing is applicable in ordinary PC platforms, the memory scale needs to be limited to 1GB, which requires $L_{\rm max}< 10$. Therefore, we believe that $L_{\rm max}=8$ could be comprehensively optimal.
		
		\section{Benchmark of matrix-operation optimization} \label{sec-performance}
		In this section, we will demonstrate the validity of our optimization with a time-complexity analysis. We focus on the overlap calculation of the even-nucleon system, i.e., Eq. (\ref{eq-ove-re}), which is the core of $M$-scheme NPA calculation.
		
		\begin{figure}
			\includegraphics[angle=0,width=0.45\textwidth]{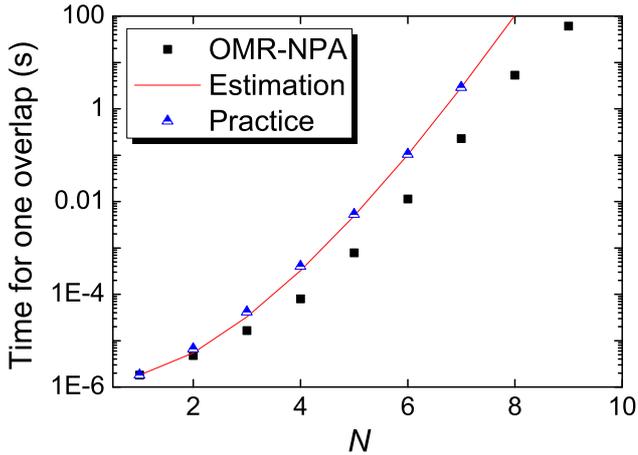}
			\caption{(color online) Computational time for one overlap calculation against nucleon-pair number $N$. ``Estimation'' represents the time estimated for the recursion calculation of Eq. (\ref{eq-ove-re}) based on Eq. (\ref{eq-tc-re}), and the assumption that one matrix multiplication or $N=1$ overlap calculation costs $1.8\times 10^{-6}$s, which is the typical time for the $pf$-shell calculation on a 4.9GHz PC platform with MKL library \cite{mkl}. ``Practice'' is the average computational time of one overlap in the basis orthogonalization of optimized NPA calculation in $pf$ shell with only $SD$ pairs, with direct programming of Eq. (\ref{eq-ove-re}) without optimization in practice. ``OMR-NPA'' corresponds the same time measure as the ``Practice" does, except with optimized OMR-NPA implementation.}\label{fig-ove-tim-en}
		\end{figure}
		
		The time-complexity analysis for Eq. (\ref{eq-ove-re}) without optimization is quite straightforward. One can consider that the $2N$-nucleon overlap is a sum of $N+N(N-1)/2$ overlaps in $2(N-1)$-nucleon system. Thus, the computational time of $2N$ overlap is approximately equal to that of $2(N-1)$ with a factor of $N(N+1)/2$. By recursively using this relation, we have
		\begin{equation}\label{eq-tc-re}
			\begin{aligned}
				T(N)&=\frac{N(N+1)}{2}T(N-1)\\
				&=\frac{N!(N+1)!}{2^N}T(1)
			\end{aligned}
		\end{equation}
		where $T(N)$ means the computational time of the unoptimized $2N$-nucleon overlap calculation. Above time-complexity regularity is presented in Fig. \ref{fig-ove-tim-en}. We also implement Eq. (\ref{eq-ove-re}) without any further optimization in practice. The practical average computational time for one overlap is compared with the time-complexity regularity of Eq. (\ref{eq-tc-re}) in Fig. \ref{fig-ove-tim-en}. The agreement between the time complexity from estimation and in practice is obvious, and thus proves the validity of Eq. (\ref{eq-tc-re}).
		
		To demonstrate the advantages of our optimization, we also plot the average computational time of OMR-NPA for one-overlap calculation in Fig. \ref{fig-ove-tim-en}, compared with the time-complexity regularity of Eq. (\ref{eq-tc-re}). One sees that our optimization indeed reduces the computational time by $1\sim 2$ magnitudes beyond the unoptimized algorithm. Such performance enhancement is more obvious for larger $N$. It is indeed worthwhile to trade-off space complexity to reduce time complexity.
		
\begin{figure}
			\includegraphics[angle=0,width=0.45\textwidth]{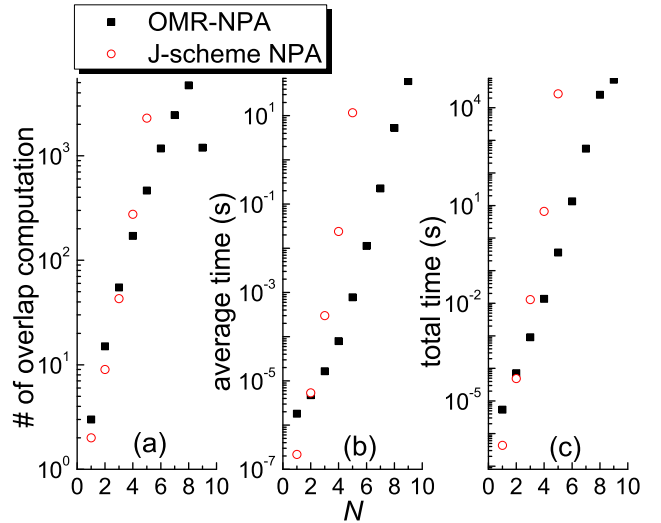}
			\caption{(color online) computational cost of basis overlap calculation against valence nucleon-pair number, $N$. The calculations are performed in $pf$ shell with $SD$ truncation. In Plane (a) we present the number of overlap calculation, (b) is for the average computational time per overlap, and (c) is for total time consumption. The $J$-scheme NPA is inapplicable for $N>5$ cases, and thus omitted here.}\label{fig-jm-com-en}
		\end{figure}

		Although the overlap calculation is improved, the actual overall improvement still may be hindered by the enlarged dimension to maintain the rotational completeness of the $M$-scheme model space. Thus, to present the actual performance of the OMR-NPA code, we calculate all the overlaps between the linearly independent base in the $pf$ shell with only the $SD$ pair, similarly to the test calculation in Sec. \ref{sec-mem-poo}. For comparison, we also calculate all the overlaps of the $J$-scheme base within the traditional NPA code. The required computational overheads in both codes are compared in Fig. \ref{fig-jm-com-en}.
		
		Fig. \ref{fig-jm-com-en}(a) shows the number of overlap calculations for different valence nucleon-pair numbers. One sees that the OMR-NPA requires more overlap calculations for $N\leq 3$ than the $J$-scheme NPA because, with angular-momentum conservation, the $J$-scheme overlap matrix is packed into several diagonal sub-matrices, where many off-diagonal overlaps are directly evaluated as zero without an overlap calculation. However, for $N>3$, the J scheme needs to deal with more overlaps. This is because the J scheme introduces the over complete intermediate quantum numbers from angular momentum recoupling, which creates more base, and thus more overlaps. The OMR-NPA does not have this over-complete issue, requires fewer overlaps for $N>3$. It is noteworthy that when $N =9$, the overlap number of OMR-NPA decreases, because the $pf$ shell is almost full for $N=9$, and thus the number of linear independent base, i.e., the overlap number, decreases.
		
		Fig. \ref{fig-jm-com-en}(b) provides a direct comparison between $J$-scheme NPA and OMR-NPA for a single overlap calculation on average. The OMR-NPA would cost less computational time than $J$-scheme by order of 1$\sim$4, which demonstrates the advantage of the matrix representation and our optimization. Such an advantage becomes more obvious as increasing the valence-nucleon pair number.
		
		Figure \ref{fig-jm-com-en}(c) comprehensively compares the total time cost of these two calculations. Since the overlap number of OMR-NPA is potentially smaller than the J scheme, and its computational time for one overlap is also less than the J scheme, its total time consumption obviously should be less than the J scheme, as shown in Fig. \ref{fig-jm-com-en}(c).

		\section{NPA-specific optimizations} \label{sec-NPA-optimization}
		Besides the optimization of the matrix operation, by invoking the Wigner-Eckart theorem, time-reversal symmetry, and conjugate property of spherical tensors, we can avoid more than half of the matrix-element calculations. In this section, we would describe these NPA-specific optimizations in detail. 
		
		In the NPA, all the  collective pairs and one-body operators to construct many-body basis, Hamiltonian, and transition operators have good angular momentum, as well as corresponding projection. Thus, the NPA matrix elements can be expressed as below, 
		\begin{equation}
			\langle 0| \mathcal P^{L_0\prime}_{M_0\prime}\cdots \mathcal P^{L_N\prime}_{M_N\prime}|O^k_{\kappa}|\mathcal P^{L_0\dagger}_{M_0}\cdots \mathcal P^{L_N\dagger}_{M_N}|0\rangle,
		\end{equation}
		where $O$ is an arbitrary spherical tensor with rank $k$, $L$ is the angular momentum of each collective pair, $M$ and $\kappa$ are the angular-momentum projections. $\mathcal P^{L_0\prime}_{M_0\prime}$ and $\mathcal P^{L_0\dagger}_{M_0}$ equal 1 or a single-particle operator following the convention of Eq. (\ref{eqn:basis}). According to the Wigner-Eckart theorem, above matrix element is non-zero, only if
		\begin{equation}
			\sum_i M_i^{\prime}\equiv \sum_i M_i+\kappa.
		\end{equation}
		Thus, there are many zero elements. According to this property, we organize the $M$-scheme basis according to the sum of angular-momentum projection, so that the matrices of Hamiltonian and transition operators can be divided into several sub-matrices, which saves memory allocation and facilitates zero-element identification.
		
		Under the time-reversal operations, the matrix element of a certain operator in the $M$-scheme NPA could be related to another matrix element, where corresponding pair and one-body operators have exact opposite angular-momentum projection as described follows. With the time-reversal operator, $\mathcal T$, any spherical tensor can be transferred as
		\begin{equation}
			\mathcal T O^k_{\kappa} \mathcal T^{-1}=(-)^{k-\kappa}O^k_{-\kappa}.
		\end{equation}
		We remind that $\mathcal P^L_M$ is not a spherical tensor, but its time-reversal operator $\tilde{\mathcal P}^L_M=(-)^{L-M}\mathcal P^L_{-M}$ is. Thus, the $\mathcal P^L_M$ follows the same transformation rule, given $\mathcal T \mathcal P^{L}_M\mathcal T^{-1}=(-)^{L+M}\mathcal T \tilde {\mathcal P}^{L}_{-M}\mathcal T^{-1}=\tilde {\mathcal P}^L_{M}=(-)^{L-M}\mathcal P^L_{-M}$. Thus, any $M$-scheme NPA matrix element could be rewritten with the time-reversal partners of involved operators as
		\begin{widetext}
			\begin{equation}\label{eq-tim}
				\begin{aligned}
					\langle 0|&\mathcal P^{L_0\prime}_{M_0\prime}\mathcal P^{L_1\prime}_{M_1\prime}\cdots \mathcal P^{L_N\prime}_{M_N\prime}|O^k_{\kappa}|\mathcal P^{L_0\dagger}_{M_0}\mathcal P^{L_1\dagger}_{M_1}\cdots \mathcal P^{L_N\dagger}_{M_N}|0\rangle\\
					&=\langle 0|\mathcal T^{-1}\mathcal T \mathcal P^{L_0\prime}_{M_0\prime}\mathcal T^{-1}\mathcal \cdots \mathcal T \mathcal P^{L_N\prime}_{M_N\prime}\mathcal T^{-1}\mathcal TO^k_{\kappa}\mathcal T^{-1}\mathcal T\mathcal P^{L_0\dagger}_{M_0}\mathcal T^{-1}\mathcal T\mathcal P^{L_1\dagger}_{M_1}\mathcal T^{-1}\cdots\mathcal T\mathcal P^{L_N\dagger}_{M_N}\mathcal T^{-1}\mathcal T|0\rangle\\
					&=(-)^{\sum_i (L_i\prime-M_i\prime+L_i-M_i)+k-\kappa}\times\langle 0| \mathcal P^{L_0\prime}_{-M_0\prime}\mathcal P^{L_1\prime}_{-M_1\prime}\cdots \mathcal P^{L_N\prime}_{-M_N\prime}|O^k_{-\kappa}|\mathcal P^{L_0\dagger}_{-M_0}\mathcal P^{L_1\dagger}_{-M_\mathcal P}\cdots \mathcal P^{L_N\dagger}_{-M_N}|0\rangle
				\end{aligned}
			\end{equation}
		\end{widetext}
		
		Under the conjugate operation, the matrix element can be related to another matrix element by exchanging the bra and ket. Especially, in the NPA, all the transition operators are hermitian or anti-hermitian spherical tensors. Thus, the conjugate operation transfers them as
		\begin{equation}
			(Q^k_{\kappa})^{\dagger}=\pm (-)^{k-\kappa}{Q}^k_{-\kappa},
		\end{equation}
		corresponding to the transpose of $q$ structural matrix. Given we are performing calculation within real number field, for all the transition-operator matrix elements, we have
		\begin{widetext}
			\begin{equation}\label{eq-con}
				\begin{aligned}
					\langle 0|\mathcal P^{L_0\prime}_{M_0\prime}\mathcal P^{L_1\prime}_{M_1\prime}\cdots \mathcal P^{L_N\prime}_{M_N\prime}|Q^k_{\kappa}|\mathcal P^{L_0\dagger}_{M_0}\cdots \mathcal P^{L_N\dagger}_{M_N}|\rangle&=\left(\langle 0 |\mathcal P^{L_0\prime}_{M_0\prime}\cdots \mathcal P^{L_N\prime}_{M_N\prime}|Q^k_{\kappa}|\mathcal P^{L_0\dagger}_{M_0}\mathcal P^{L_1\dagger}_{M_1}\cdots \mathcal P^{L_N\dagger}_{M_N}|0\rangle\right)^{\dagger}\\
					&=\pm (-)^{k-\kappa}\langle 0|\mathcal P^{L_0}_{M_0}\cdots \mathcal P^{L_N}_{M_N}|Q^k_{-\kappa}|\mathcal P^{L_0\prime\dagger}_{M_0\prime}\cdots \mathcal P^{L_N\prime\dagger}_{M_N\prime}|0\rangle.
				\end{aligned}
			\end{equation}
		\end{widetext}
		For Hamiltonian matrix element and overlap calculations, the above transformation is reduced to a sample exchange of bra and ket without any phase factor.
		
		Using Eqs. (\ref{eq-tim}) and (\ref{eq-con}), any matrix element of Hamiltonian and transition operators, as well as overlap, can be related to three other elements. Namely, we can calculate one of them, and the other three elements are obtained with a certain phase factor. Thus, 75\% of non-zero matrix-element calculations can be omitted. After all optimizations in this section and Sec. \ref{sec-opt}, the OMR-NPA approximately doubles the NPA applicable region on the nuclear chart as demonstrated in Ref. \cite{mnpa-lei}. 
		
		\section{summary} \label{sec-summary}
		
		For $M$-scheme NPA calculations, a formalism based on the matrix representation of collective pair structure was proposed \cite{mnpa-lei}, where the major computational bottleneck is attributed to the matrix operations. Therefore, we optimize the matrix operations, to gain higher computational efficiency. The corresponding implementation is denoted by OMR-NPA. In such an implementation, we express the structural coefficient matrix of any newly generated collective pair as a linear combination of several products of artificially inputted matrix by using unsigned-char arrays and accordingly store the intermediate matrix products orderly in the memory for further retrieval. Some other optimizations related to recursion, symmetric matrix multiplication, trace calculations, and properties of the NPA matrix elements are also introduced. All these optimizations efficiently reduce the time complexity of overlap and Hamiltonian matrix element calculation, which has been confirmed from various aspects as shown in this work and Ref. \cite{mnpa-lei}. 
		
		We note that the optimization described in Sec. \ref{sec-opt} is not restricted to the OMR-NPA calculations, as it is also valid for other particle-number-conserved many-body calculations, only if the matrix representation is adopted, e.g., in the recently proposed pair condensate variation \cite{pcv-lei}. We hope this work can be beneficial for nuclear physicists, who may need the NPA to analyze the low-lying structure of nuclei, and those, who wish to implement and optimize the matrix representation by themselves.
		
		Finally, our optimization frequently involves skew-symmetric matrix multiplications. If the industry could provide an optimized API for such multiplications, it can be a great boost for the matrix-represented modeling for nuclear structure.
		
		\section*{Acknowledgements}
		Y. Lei is grateful for the financial support of the Sichuan Science and Technology Program (Grant No. 2019JDRC0017), the Doctoral Program of Southwest University of Science and Technology (Grant No. 18zx7147).
		
		Y. Lu acknowledges support from the National Natural Science Foundation of China (11705100), Higher Educational Youth Innovation Science and Technology Program Shandong Province (2020KJJ004), and Taishan Scholar Project of Shandong Province (Grant No. tsqn202103062).
		
		
		
		
		\bibliographystyle{elsarticle-num}
		\bibliography{myref}
		
		
		
		
		
		
	\end{CJK}
\end{document}